\documentstyle [12pt] {article}
\title{BLACK HOLES - A SIMPLIFIED THEORY FOR QUANTUM GRAVITY NON-SPECIALISTS }

\author{Vladan Pankovi\'c\\
Department of Physics, Faculty of Sciences, 21000 Novi Sad,\\ Trg
Dositeja Obradovi\'ca 4. , Serbia, vpankovic@if.ns.ac.yu}

\date {}
\begin{document}
\maketitle
\vspace {0.5cm}

 \vspace {1.5cm}

\begin {abstract}
In this work we present a simplified description and calculation
of the Kerr-Newman black hole basic dynamical (horizons) and
thermo-dynamical (Bekenstein-Hawking entropy, Bekenstein
entropy/surface quantization, Hawking temperature and radiation)
characteristics. Also, a possibility of the fission of nearly
extremal black hole is considered in full analogy with remarkable,
simple Bohr-Wheeler theory of the nuclear fission. Given black is
physically based on the well-known principles of the classical
physics (mechanics, thermodynamics and electro-dynamics). It
includes the non-relativistic quantum mechanics and statistical
mechanics too. Finally, it includes the elementary form of the
general relativistic equivalence principle only. Also, suggested
simplified description includes mathematically, practically, only
simple algebraic equations. Here many steps are extremely
simplified and formal, representing, in fact, a linear
approximation of the quantum gravity theories (black hole entropy
and temperature can be obtained by ground standing wave at
horizon, while Hawking radiation can be obtained by  gravitational
decay of this standing wave in its traveling components). But, the
final results, i.e. formulas on the black hole basic
characteristics, are mostly effectively identical to corresponding
results obtained by accurate quantum gravity theories. In this way
suggested description can be very useful for the quantum gravity
non-specialists.
\end {abstract}

\section {Introduction}
As it is well-known black hole represents one of the most
fascinating physical object. But accurate analysis of the
dynamical and thermo-dynamical characteristics of the black hole
needs knowledge of the subtle details of general theory of
relativity and quantum field theory, i.e. quantum gravity (even if
there is no complete theory of the quantum gravity to this day)
[1]-[7]. For this reason non-specialists (that do no know very
much on the quantum gravity) cannot realize any useful calculation
of the basic dynamical (horizons) and thermo-dynamical
(Bekenstein-Hawking entropy, Bekenstein entropy/surface
quantization, Hawking temperature and radiation) characteristics
of the black hole. Of course, there are some very interesting
presentations of the black hole basic characteristics [8] for the
non-specialists but without a detailed calculation tools. In some
other references, with indicative titles, e.g. Can One Understand
Black Hole Entropy without Knowing Much about Quantum Gravity [9],
calculation tools are too complicated for the non-specialists.

In this work we shall present a simplified method for description
and calculation of the Kerr-Newman black hole main dynamical and
thermo-dynamical characteristics. This method is physically based
on the well-known principles of the classical physics (mechanics,
thermodynamics and electro-dynamics). It includes the
non-relativistic quantum mechanics and statistical mechanics too.
Finally, it includes the elementary form of the general
relativistic equivalence principle only. Also, suggested
simplified description includes mathematically, practically, only
simple algebraic equations. Here many steps are extremely
simplified and formal, representing, in fact, a linear
approximation of the quantum gravity theories (black hole entropy
and temperature can be obtained by ground standing wave at
horizon, while Hawking radiation can be obtained by  gravitational
decay of this standing wave in its traveling components). But, the
final results, i.e. formulas on the black hole basic
characteristics, are mostly effectively identical to corresponding
results obtained by accurate quantum gravity theories. In this way
suggested description can be very useful for the quantum gravity
non-specialists.

\section {Black hole dynamics}

As it is well-known [10], [11] Laplace determined by simple,
classical mechanical method, the radius $R$ of a static black
star, i.e. a star with sufficiently large mass $M$ (homogeneously
distributed over volume) so that even light cannot escape from the
star surface. He defined the total energy of a probe particle with
mass $m$ that propagates with speed of the light $c$ radially,
i.e. perpendicularly to the star surface. This total energy
$E_{tot}$ represents the sum of the probe particle classical
translational kinetic energy $\frac{mc^{2}}{2}$ and negative
potential energy of the gravitational attractive interaction
between probe particle and black star $- \frac {GmM}{R}$ were $G$
represents the Newtonian gravitational constant. Laplace supposed
that given total energy equals zero, i.e.
\begin {equation}
         E_{tot}= \frac{mc^{2}}{2} - \frac {GmM}{R} = 0         .
\end {equation}
It represents a simple algebraic equation, called Laplace
equation, with $R$ as unknown variable. In the natural system of
the units (where $G=c=1$) equation (1) turns out in
\begin {equation}
         E_{tot} = \frac {m}{2} - \frac {mM}{R}= 0
\end {equation}
with corresponding unique solution
\begin {equation}
         R = 2M           .
\end {equation}

Surprisingly given radius is identical to the Schwarzschild radius
of corresponding Schwarzschild (static, non-charged) black hole
predicted accurately by general theory of relativity [2].

In this way presented Laplace's method (equation) can be
considered as an extremely simplified method for determination of
the Schwarzschild black hole horizon as the basic dynamical
characteristic of the Schwarzschild black hole.

Laplace's method (equation) can be simply generalized in the
situation when black star is electrically charged (with charge $Q$
homogeneously distributed over black star surface) and when it
rotates (with angular momentum $J$ around z-axis).

Suppose again that total energy of the probe particle equals zero,
i.e.
\begin {equation}
 E_{tot} = \frac {mc^{2}}{2} + \frac {m}{2}(\frac {c{\it a}}{R})^{2} - \frac {GmM}{R} + Gm[\frac {1}{2}\frac {Q^{2}}{4\pi \epsilon_{0}Rc^{2}}]\frac {1}{R} = 0  .
\end {equation}
It again represents a simple algebraic equation, called
generalized Laplace equation, with $R$ as the unknown variable.

Here $\frac {mv^{2}}{2}= \frac {m}{2}(\frac {c{\it a}}{R})^{2}$
represents the classical kinetic energy of the rotation of probe
particle at black star surface with peripheral speed $v$ induced
by black star rotation. It includes the relation $mvR=mc{\it a}$
and relation $J=Mc{\it a}$. First relation simply means that ${\it
a}$ represents formally a distance at which probe particle would
rotate with peripheral speed $c$. Then corresponding formal
angular momentum $mc{\it a}$ is equivalent to the real angular
momentum $mvR$. Second relation is not simply and it represents
the exact general relativistic definition of ${\it a}$ as distance
parameter for rotating black hole with mass $M$ and angular
momentum $J$. Namely, classical mechanical rigid body, with radius
$R$ and homogeneously distributed over volume mass $M$, holds
momentum of the inertia $\frac {2}{5}MR^{2}$ but not $MR^{2}$. It
implies classical angular momentum $\frac {2}{5}MvR=\frac
{2}{5}Mc{\it a}$ but not $MvR= Mc{\it a}$. In this way use of $
Mc{\it a}$  instead of $ \frac {2}{5}Mc{\it a}$   in (4)
represents an ad hoc postulated correction of the classical
expression. It is, in some way, similar to situation by definition
of the electron mass and radius in the classical electrodynamics
[12].

Further, $V_{e}=\frac {1}{2}\frac {Q^{2}}{4\pi \epsilon_{0}R}$
represents the classical potential energy of the electrostatic
Coulomb's repulsion for the sphere with radius $R$ and electrical
charge $Q$ homogeneously distributed over sphere surface [12]
where $\epsilon_{0}$ represents the vacuum dielectric
permeability. Also, according to general relativistic equivalence
principle, $|M_{e}|=\frac {V_{e}}{c^{2}}$ can be absolute value of
the mass corresponding to $V_{e}$, i.e. to electrostatic mass.
But, since $V_{e}$ corresponds to a repulsive interaction it can
be supposed that electrostatic mass $M_{e}$ is negative, i.e. that
$M_{e}= -\frac {V_{e}}{c^{2}}$. Finally, $ Gm[\frac {1}{2}\frac
{Q^{2}}{4\pi \epsilon_{0}Rc^{2}}]\frac {1}{R} = -\frac {Gm
M_{e}}{R} $ can be formally considered as the positive potential
energy of the classical gravitational interaction between
(negative) electrostatic mass and (positive) probe particle mass.

In this way total energy of the probe particle is presented as the
sum of the total (translational and rotational) classical kinetic
energy and total (formally) classical gravitational (basic
gravitational and electrostatic presented as the gravitational by
equivalence principle) potential energy.

In the natural system of units (where $4\pi \epsilon_{0}=1$)
equation (4) turns out in
\begin {equation}
 E_{tot} =  \frac {m}{2} + \frac {m}{2}(\frac {{\it a}}{R})^{2} - \frac {mM}{R} + m\frac {1}{2}\frac {Q^{2}}{R^{2}}= 0
\end {equation}
or, after simple transformations, in
\begin {equation}
 R^{2} -2MR + ({\it a}^{2} + Q^{2})= 0   .
\end {equation}
This equation, generally speaking, holds two solutions
\begin {equation}
 R_{\pm}= M \pm (M^{2} - ({\it a}^{2} + Q^{2}))^{\frac {1}{2}}           .
 \end {equation}

Surprisingly $R_{+}$ and $R_{-}$ are identical to the outer and
inner horizon of the Kerr-Newman (rotating with angular momentum
$J$, i.e. distance parameter {\it a}, and electrically charged
with charge $Q$) black hole predicted accurately by general theory
of relativity [2].

But, given solutions (7) are real and different only for
$M^{2}\geq ({\it a}^{2} + Q^{2})$, when Kerr-Newman black hole is
called non-extremal.

For $M^{2}= ({\it a}^{2} + Q^{2})$, when Kerr-Newman black hole is
called extremal, (7) represents unique real solution of (5).

Finally, for $M^{2}< ({\it a}^{2} + Q^{2})$, when Kerr-Newman
black hole is called super-extremal, equation (5) does not hold
real solutions which implies black star non-existence. Namely, in
this case $E_{tot}$ is always positive so that light can always
escape from star surface. It is, in some way, in agreement with
accurate prediction of the general relativity according to which
condition $M^{2}< ({\it a}^{2} + Q^{2})$ implies implausible
breaking of the cosmic censorship theorem, i.e. appearance of the
naked singularities.

Generalized Laplace equation (5), representing formally
equilibrium condition for probe particle, can be, in fact,
considered as the condition of the dynamical stability of whole
Kerr-Newman black hole. It can be roughly demonstrated and
described by classical physics.

Namely, non-extremal Kerr-Newman black hole is, roughly
classically speaking, dynamically stable since here attractive,
i.e. centripetal, gravitational force is much larger than sum of
two repulsive, forces, electrostatic and rotational, i.e.
centrifugal.

Vice versa, super-extremal Kerr-Newman black hole is, roughly
classically speaking, dynamically extremely non-stable since here
attractive, i.e. centripetal, gravitational force is much weaker
than sum of two repulsive, forces, electrostatic and rotational,
i.e. centrifugal. Then it decays or it does a spontaneous fission
practically instantaneously in the extremely many fragments of the
usual, non-collapsing systems.

Extremal Kerr-Newman black hole is, roughly classically speaking,
dynamically meta-stable since here attractive, i.e. centripetal,
gravitational force is equal to sum of two repulsive, forces,
electrostatic and rotational, i.e. centrifugal. It implies that a
non-extremal, but near to extremal, Kerr-Newman black hole,
stimulated by absorption of an additional, activating system (with
relatively small mass and relatively large charge and angular
momentum), can decay or can do a stimulated fission in two
approximately equivalent, strictly non-extremal, Kerr-Newman black
holes. Precisely, it can be supposed that, after fission, first,
1, and second, 2, fission fragment hold the same electrical charge
$q_{1}=q_{2}=\frac {Q}{2}$ and the same angular momentums, i.e.
distance parameters ${\it a}_{1}={\it a}_{2}=\frac {{\it a}}{2}$,
according to electrical charge and orbital momentum conservation
laws. It can be supposed too that both fission fragments hold the
same masses $m_{1}=m_{2}=m$ and outer horizons
$r_{1}=r_{2}=m+(m^{2} - ((\frac {{\it a}}{2})^{2}+(\frac
{Q}{2})^{2}))^{\frac {1}{2}}$. Given masses can be determined by
condition $M = m_{1}+ m_{2}- \frac {m_{1}m_{2}}{r_{1}+ r_{2}}$. In
this condition last term, i.e. potential of the classical
gravitational attractive interaction between fragments captured in
the initial extremal black hole, can be considered as a typical
relativistic mass excess (defect) Given condition represents the
algebraic quadratic equation. It holds simple positive solution
$m\simeq 1.45 M$. It implies $m^{2}\simeq 2M^{2}$ sufficiently
larger than $\frac {{\it a}}{2})^{2}+(\frac {Q}{2})^{2} = \frac
{M^{2}}{4}$ which means that fission fragments are really
non-extremal. (Obviously, described nearly extremal Kerr-Newman
black hole fission is in a deep conceptual analogy with remarkable
Bohr-Wheeler simple, approximate, quasi-classical model of the
nuclear fission.)

Finally, it can be observed that equation (6), for outer horizon
(in further text outer horizon $R_{+}$  will be denoted $R$, i.e.
without subscript $+$, for reason of simplicity), it follows
\begin {equation}
  M = \frac {R}{2} + \frac {{\it a}^{2}}{2R} +  \frac {Q^{2}}{2R}          .
\end {equation}
Term $\frac {R}{2}$ in (8) corresponds to the Newtonian, static
gravitational mass. Term $\frac {{\it a}^{2}}{2R}$ in (8) can be
treated as the rotational, or according to equivalence principle,
local gravitational (induced by rotation) mass. Then term $M_{g} =
\frac {R}{2}+  \frac {{\it a}^{2}}{2R}$ can be considered as the
pure (static and rotational) gravitational mass of the black star,
i.e. black hole. Term $\frac {Q^{2}}{2R}$ can be considered as the
electrostatic mass of the black star, i.e. black hole.

In this way presented generalized Laplace's method can be
considered as an extremely simplified method for determination of
the basic dynamical characteristics of the Kerr-Newman black hole.

\section {Black hole thermodynamics}

Suppose now that the gravitational mass of the black hole $M_{g}$
is quantized and that its quantums satisfy the following
quantization condition
\begin {equation}
      m_{n}cR = n\frac {\hbar}{2\pi}, \hspace{1cm} {\rm for} \hspace{0.5 cm}
    n=1,2,3,...
\end {equation}
that implies
\begin {equation}
      2\pi R = n\frac {\hbar}{m_{n}c}= n \lambda_{rn}  \hspace{1cm} {\rm for} \hspace{0.5 cm}
    n=1,2,3,...
\end {equation}
where $\hbar$ represents the reduced Planck's constant while
$\lambda_{rn}=\frac {\hbar}{m_{n}c}$, represents reduced Compton's
wave length corresponding to $m_{n}$ for $n=1,2, …$ . Last two
expressions, in natural units system (where $\hbar=1$) turns out
in
\begin {equation}
      m_{n} R = n\frac {1}{2\pi}       ,       \hspace{1cm} {\rm for} \hspace{0.5 cm}
    n=1,2,3,...
\end {equation}
that implies
\begin {equation}
      2\pi R = n \frac {1}{m_{n}} = n \lambda_{rn}    \hspace{1cm} {\rm for} \hspace{0.5 cm}
    n=1,2,3,...        .
\end {equation}
Last expression means, in fact, that the "circumference" of the
outer horizon (even if, of course, outer horizon surface does not
represent exactly a sphere) holds $n$ reduced Compton's wavelength
of the mass quantums with mass $m_{n}$ for $n = 1,2,...$ .

Obviously, (9), (10) or (11), (12) correspond, in some degree, to
Bohr's postulate on the electron orbital momentum quantization and
de Broglie's interpretation of this postulate in the atomic
physics. (However, electron orbit radius represents a variable
that increases proportionally to $n^{2}$ while all gravitational
mass quantums of the black hole, hold the same orbit radius $R$.
Also, electron energy represents a variable that increases
proportionally to $-\frac {1}{n^{2}}$, while quantum of the black
hole gravitational mass, represents a variable that increases
proportionally to $n$, i.e. linearly.)

Expression (11) implies
\begin {equation}
       m_{n} = n \frac {1}{2\pi R} = n m_{1}    \hspace{1cm} {\rm for} \hspace{0.5 cm}
    n=1,2,3,...
\end {equation}
where
\begin {equation}
       m_{1} = \frac {1}{2\pi R}
\end {equation}
represents the minimal, i.e. ground mass of the gravitational mass
quantums.

Further, suppose that black hole gravitational mass quantums do a
statistical ensemble. In other words, suppose that there is a
gravitational self-interaction of the black star, i.e. black hole
which can be described statistically. Suppose that in the
thermodynamical equilibrium almost all quantums occupy ground mass
state. It implies that black hole gravitational mass quantums
represent the Bose-Einstein quantum systems, i.e. bosons. Then,
roughly speaking, corresponding entropy $S$ can be defined by
\begin {equation}
  S = k_{B}\frac {M_{g}}{m_{1}}
\end {equation}
where $k_{B}$ represents the Boltzmann constant, or, in the
natural system units (where $ k_{B}=1$)
\begin {equation}
  S = \frac {M_{g}}{m_{1}}=  \pi (R^{2} + {\it a}^{2}) = \frac {A}{4}
\end {equation}
where $A= 4\pi(R^{2} + {\it a}^{2})$ represents the black star,
i.e. hole outer horizon surface.

Surprisingly, $S$ (16) is identical to the Bekenstein-Hawking
entropy of the Kerr-Newman black hole obtained by the accurate
quantum gravity methods [1]-[4].

Suppose that a is much smaller than $R$ and $M$ and then
differentiate $S$ (16) over $R$ that yields
\begin {equation}
    dS = 2\pi RdR                .
\end {equation}

Further, according to (7), for a as well as $Q$ much smaller than
$M$, it follows
\begin {equation}
    dR = (1 + \frac {M}{(M^{2} - {\it a}^{2} - Q^{2})^{\frac {1}{2}}}) dM = \frac {R}{(M^{2} - {\it a}^{2} - Q^{2})^{\frac {1}{2}} }dM
\end {equation}
or, in more rough approximation,
\begin {equation}
    dR \simeq 2dM        .
\end {equation}
Then (17), (18), (19) imply
\begin {equation}
    dS = 2 \pi \frac {R^{2}}{(M^{2} - {\it a}^{2} - Q^{2})^{\frac {1}{2}} }dM
\end {equation}
or, in more rough approximation,
\begin {equation}
   dS \simeq 4\pi RdM
\end {equation}

Introduction of (20), in (17), yields
\begin {equation}
    dS \simeq 4\pi R dM
\end {equation}
or, in the additional approximation, i.e. by change of the
differentials by finite differences,
\begin {equation}
    \Delta S \simeq 4\pi R \Delta M    \hspace{1cm} {\rm for} \hspace{0.5 cm}   \Delta M \ll M  .
\end {equation}
Further, we shall assume
\begin {equation}
   \Delta M = n m_{1} \hspace{1cm} {\rm for} \hspace{0.5 cm}
    n=1,2,3,...    \hspace{1cm}   {\rm and} \hspace{0.2cm} {\rm for} \hspace{0.5cm}   \Delta M \ll M   .
\end {equation}
It, introduced in (23) and according to (14), (16), yields
\begin {equation}
    \Delta S = 2n  \hspace{1cm} {\rm for} \hspace{0.5 cm}
    n=1,2,3,...    \hspace{1cm}   {\rm and} \hspace{0.2cm} {\rm for} \hspace{0.5cm}   \Delta M \ll M
\end {equation}
and
\begin {equation}
    \Delta A = 2n (2)^{2}  \hspace{1cm} {\rm for} \hspace{0.5 cm}
    n=1,2,3,...    \hspace{1cm}   {\rm and} \hspace{0.2cm} {\rm for} \hspace{0.5cm}   \Delta M \ll M
\end {equation}
where 2 can be considered as the twice Planck length in the
natural units system.

Surprisingly, $\Delta S$ (25) and $\Delta A$ (26) are identical to
the Bekenstein's quantization of the black hole entropy and
horizon surface area.

All this implies that black star, i.e. black hole can be
considered as a thermodynamical system in the state of the
thermodynamical equilibrium. Then the first thermodynamical law,
in the following form,
\begin {equation}
   dM = T dS + \Omega dJ + \Phi dQ
\end {equation}
must be satisfied, where
\begin {equation}
    \Omega = \frac {{\it a}}{R^{2}+ {\it a}^{2}}
\end {equation}
represents the outer horizon angular speed. Also, according to
accurate quantum gravity theories $\Phi= \frac {QR}{ R^{2}+ {\it
a}^{2}}$ represents the outer horizon electrostatic potential. As
it is not hard to see $\Phi$ is different from previously
implicitly defined electrostatic potential $\frac {V_{e}}{Q}$.
But, as it will be demonstrated, given difference does not hold
any important role in the further thermodynamical considerations.

Further, we shall approximately neglect term $\Phi dQ$ in (27) so
that this expression, according to (28), turns out in
\begin {equation}
  dM = TdS +  \Omega dJ = TdS +  \frac {{\it a}}{R^{2}+ {\it a}^{2}} dJ .
\end {equation}
This expression, according to $J={\it a}M$ definition,
approximation condition ${\it a} \ll  M$, and (20), implies
\begin {equation}
  dM= T 2 \pi \frac {R^{2}}{(M^{2}- {\it a}{2}- Q^{2})^{\frac {1}{2}}}dM + \frac {{\it a}^{2}}{R^{2}+ {\it a}^{2}}dM.
\end {equation}
equivalent to the following simple algebraic equation with $T$ as
single unknown variable
\begin {equation}
  1 = T 2\pi \frac {R^{2}}{(M^{2}- {\it a}{2}- Q^{2})^{\frac {1}{2}}}+ \frac {{\it a}^{2}}{R^{2}+ {\it a}^{2}}         .
\end {equation}
Solution of the last equation is simply
\begin {equation}
  T  = \frac {1}{2\pi}\frac { (M^{2}- {\it a}{2}- Q^{2})^{\frac {1}{2}}}{ R^{2}+ {\it a}^{2}}           .
\end {equation}

Surprisingly $T$ (32) is identical to the Hawking temperature of
the black hole obtained by accurate methods of the quantum gravity
[1]-[4].

In this way presented method can be considered as an extremely
simplified method for determination of the basic thermodynamical
characteristics of the Kerr-Newman black hole.

\section {Black hole statistical mechanics}

Now, we shall give a deeper, statistical interpretation of the
black hole entropy.

It has been supposed previously that black hole gravitational mass
quantums do a bosonic great canonical  ensemble in the
thermodynamical equilibrium, with mass spectrum $m_{n}$ for $n =
1,2,...$ (13), temperature $T$ (32) and chemical potential $\mu$
whose value will be determined later.

Then, as it is well-known and according to (13), statistically
averaged number of the quantums with mass $m_{n}$, $N_{n}$, for $n
= 1,2,... $, is given by expression
\begin {equation}
 N_{n} = \frac {g_{n}}{\exp[\frac {m_{n} - \mu}{T}] -1} = g_{n} /(exp[(\frac {nm_{1} - \mu}{T}] -1) \hspace{1cm} {\rm for} \hspace{0.5 cm}
    n=1,2,3,...
\end {equation}
where $ g_{n}$ represents the degeneracy of the quantum state
corresponding to $m_{n}$  for $n = 1,2,... $.

Also, as it is well-known too, partial entropy in the quantum
state corresponding to $m_{n}$  for $n = 1,2,... $ , is given by
expression
\begin {equation}
 S_{n} = g_{n} \ln [1 + \frac {N_{n}}{g_{n}}] + N_{n} \ln [1 + \frac {g_{n}}{N_{n}}]  \hspace{1cm} {\rm for} \hspace{0.5 cm}
    n=1,2,3,...
\end {equation}
where $g_{n}$ represents the degeneracy of the quantum state
corresponding to $m_{n}$  for $n = 1,2,... $.

We shall suppose
\begin {equation}
   g_{n} \simeq 1  \hspace{1cm} {\rm for} \hspace{0.5 cm}       n \gg 1
\end {equation}
which, according to (33), (34) implies
\begin {equation}
 N_{n}\ll 1     \hspace{1cm} {\rm for} \hspace{0.5 cm}   n \gg 1
\end {equation}
and
\begin {equation}
 S_{n} \simeq N_{n}\ll 1   \hspace{1cm} {\rm for} \hspace{0.5 cm}    n \gg 1              .
\end {equation}

Also, we shall suppose
\begin {equation}
 g_{1} = N_{1}                                                   .
\end {equation}
It, according (14), (32), (33) implies the following value of the
chemical potential
\begin {equation}
  \mu = m_{1} - T \ln2 = m_{1}(1 - \frac {T}{m_{1}}\ln2 ) .
\end {equation}

Intuitive explanation of the suppositions (35), (38) is very
simple. Ground mass state corresponding to $m_{1}$,
(energetically) closest to (outer) horizon, maximally exposed to
gravitational influence, is maximally degenerate. Highly excited
quantum states corresponding to $m_{n}$ for $n \gg 1$,
(energetically) very distant from horizon, are not so strongly
exposed to gravitational influence and are almost non-degenerate.

It can be observed that here we have a situation in some degree
similar to Bose condensation. Black hole gravitational mass
quantums occupy maximally, maximally degenerate ground mass state,
in respect to other, practically non-degenerate, mass states (even
if, according to (38), $\frac {N_{1}}{g_{1}}$ does not tend toward
infinity but toward 1).

According to (34)-(38) it follows
\begin {equation}
  S_{1} = 2 \ln2 N_{1}\simeq 1.39 N_{1} \sim N_{1}    \hspace{1cm} {\rm for} \hspace{0.5 cm}      n = 1               .
\end {equation}
It implies the following expression for usually statistically
defined total entropy $S$
\begin {equation}
  S = \sum_{n=1}  S_{n}\simeq S_{1}\simeq 1.39 N_{1}\sim N_{1}
\end {equation}
and equivalence of  (41) and (40) implies
\begin {equation}
  N_{1}\simeq \frac {1}{1.39}\frac {M_{g}}{m_{1}}\simeq 0.72  \frac {M_{g}}{m_{1}}\sim  \frac {M_{g}}{m_{1}}            .
\end {equation}
Then, statistically averaged total number of the mass quantums N
is given by expression
\begin {equation}
  N = \sum_{n=1}N_{n}\simeq N_{1} \simeq \frac {1}{1.39}\frac {M_{g}}{m_{1}}\simeq 0.72  \frac {M_{g}}{m_{1}}\sim  \frac {M_{g}}{m_{1}}   .
\end {equation}
Statistically averaged black hole gravitational mass <Mg> is given
by expression
\begin {equation}
  <M_{g}> = \sum_{n=1}N_{n} m_{n}\simeq N_{1}m_{1}\simeq \frac {1}{1.39}M_{g}\simeq 0.72 M_{g}\sim  M_{g}
\end {equation}
which corresponds approximately to black hole gravitational mass
$M_{g}$.

Thus, black hole entropy is practically equivalent to the number
of the black hole gravitational mass quantums in the ground state
or to the degeneracy of given state.

It can be observed that accurate quantum gravity theories predict
the canonical statistical ensemble of the quantum systems emited
by Hawking radiation. Within given accurate theories
micro-canonical distribution is inconsistent (it yields
non-plausible statements), while macro-canonical distribution is
superfluous. Within suggested approximate theory, as it is not
hard to see, both, micro-canonical and canonical distribution are
inconsistent and only great-canonical distribution, with
especially chosen chemical potential (39), can effectively
reproduce expected thermodynamical variables.

In this way it is demonstrated that previously discussed basic
thermodynamical characteristics of the black hole can be founded
statistically in a simple, but satisfactory approximation.

\section {Hawking radiation}

Hawking radiation and evaporation of the black hole can be usually
interpreted in the following way (for details see [1], [3]).
Quantum fluctuations of the vacuum and corresponding to Heisenberg
energy-time uncertainty relation, that appear outside but nearly
to black hole (outer) horizon, imply creation of a virtual pair of
the particle and anti-particle. Suppose that one member of the
pair propagates toward horizon and falls in the black hole. For a
distant observer in the asymptotically flat coordinate system
energy of the infalling member is negative. Other member in the
pair propagates in the opposite direction, i.e. far away from
horizon toward mentioned distant observer. For this distant
observer given member seems effectively as the radiation emited by
black hole. More accurately, Hawking showed that emited particles
satisfy canonical Bose-Einstein statistical distribution without
degeneration, or that black hole behaves as a black body that
satisfies Planck radiation law at Hawking temperature. It implies
(after well-known statistical physical formalism that includes
integration over all frequencies of the radiation etc.) that total
energy emited over black hole horizon surface per time unit can be
described by Stefan-Boltzmann law
\begin {equation}
  -\frac {dM}{dt} c^{2} = \sigma T^{4} A
\end {equation}
where $\sigma= \frac {\pi^{2}k^{4}_{B}}{60\hbar^{3}c^{2}}$, or
$\sigma= \frac {\pi^{2}}{60}$   in the natural system units,
represents the Stefan-Boltzmann constant. It yields, according to
(7), (16) and (32), after simple integration, the time of the
total evaporation of the black hole $t_{ev}$, corresponding to the
decrease of the initial black hole mass $M$ to zero.

In the simplest case, i.e. for Schwarzschild black hole, and in
the natural system units, Stefan-Boltzmann law (45) turns out in
\begin {equation}
   -\frac {dM}{dt}  = \frac {1}{15360 \pi  M^{2}}
\end {equation}
that, after simple integration, yields
\begin {equation}
    t_{ev} = (5120 \pi)M^{3}                   .
\end {equation}
(Similar procedure can be realized in the general case, i.e. for
Kerr-Newman black hole, which here will not be realized
explicitly, for reason of the simplicity.)

Now, it will be attempted that an effective correspondence be
realized between mentioned accurate theory of the Hawking
radiation and a simplified theory of the Hawking radiation.

It can be supposed that black hole gravitational mass quantums
$m_{n}$ linearly dependent of $n$ for $n = 1,2,... $ (13)
correspond to simple, sinusoidal (corresponding to a linear
dynamics) standing waves with wave lengths $\lambda_{rn}= \frac
{1}{m_{n}}$  for $n = 1,2,...$  (12) along "circular" string $2\pi
R$ equivalent to horizon "circumference". (Word string is used
here in the sense of the usual, classical mechanics, but not in
the sense of the string theories of quantum gravity.) Any of given
standing waves can be considered as the usual superposition of two
simple, sinusoidal traveling waves. They propagate in the opposite
directions, or with opposite circular frequencies
$\omega_{n}=m_{n}$  and $-\omega_{n}=-m_{n}$  for $n = 1,2,...$
and with one half of the amplitude of corresponding standing wave
over horizon "circumference". More accurately, i.e. from
relativistic quantum theory view point, one of the traveling wave
can be considered as the particle, other - as the anti-particle.
Simultaneously, corresponding standing wave can be considered as
the typical zitterbewegung (trembling motion), i.e. superposition
of the particle and anti-particle.

It can be supposed that given gravitational mass quantums, i.e.
corresponding standing waves are not dynamically completely
stable. Namely, given mass quantums represent quantums of the
black hole gravitational self-interaction too. But an additional
small perturbation, corresponding to gravitational interaction
between given quantums and black hole, can be supposed too. For
this reason, it can be supposed that any gravitational mass
quantum $m_{n}$ can decay into its two traveling waves components,
i.e. particle and anti-particle, after some (statistically
averaged) life time interval $\tau_{n}$ for $n = 1,2,...$ . It can
be supposed too that, by decay, one of the traveling waves falls
inside horizon while other traveling waves goes far away from
horizon toward a distant observer in the asymptotically flat
coordinate system.

Obviously, all this corresponds in some degree to mentioned,
accurate description of the Hawking radiation.

But, it seems that there is a principal distinction between
mentioned accurate theory of the Hawking radiation and simplified
theory of the Hawking radiation. As it has been discussed and
pointed out accurate theory considers canonical Bose-Einstein
distribution without degeneration of the particles with all
possible frequencies from zero till infinity. Simplified method
considers, as it has been demonstrated in the previous section of
this work, macro-canonical Bose-Einstein distribution with
occupied practically only degenerate, ground, non-zero mass state.
In other words, practically only black hole gravitational mass
quantum in the degenerate ground state corresponding to m1 is
important in the suggested, simplified description of the black
hole radiation. Or, in given simplified description of black hole
radiation practically only one standing wave with unique wave
length $\lambda_{r1}=\frac {1}{m_{1}}$ and life time $\tau_{1}$
and corresponding traveling waves (i.e. particle and anti-particle
with circular frequencies $\omega_{1}=m_{1}$  and
$-\omega_{1}=-m_{1}$  and (statistically averaged) amplitudes
twice smaller than amplitude of the standing wave) can have
important role. For this reason here Stefan-Boltzmann law cannot
be obtained.

Nevertheless, suppose that an effective correspondence between
usual and simplified description of Hawking radiation can be
realized in the simplest case, i.e. for Schwarzschild black hole,
in the following way.

Suppose that accurate, differential form of the black hole mass
decrease in the time unit -$\frac {dM}{dt}$ in the
Stefan-Boltzmann law (46) can be approximately changed, i.e.
linearized by the finite differences quotient $\frac
{m_{1}}{\tau_{1}}$. It yields
\begin {equation}
   \frac {m_{1}}{\tau_{1}} = \frac {1}{15360 \pi  M^{2}}
\end {equation}
that, according to (14), (43), implies
\begin {equation}
   \tau_{1} = 15360 \pi m_{1}M^{2}= 3840 M       .
\end {equation}
So, for life time of the ground mass quantum determined by (49)
accurate, differential (Stefan-Boltzmann) and approximate, finite
black hole mass decrease will be effectively the same.

From (48) it follows
\begin {equation}
   \frac {1}{\tau_{1}} = \frac {1}{15360 \pi m_{1}M^{2}} = 2\pi |V_{p}|^{2}\rho              .
\end {equation}
Here $V_{p}= -\frac {m_{1}M}{R} = \frac {1}{8\pi M} = - T$
represents the small perturbation corresponding to classical
gravitational interaction between ground mass quantums and black
hole. It is very interesting that such perturbation corresponds to
negative Hawking temperature. Also, $\rho=\frac {1}{480m_{1}}=
\frac {4\pi M}{480} $ can be phenomenologically interpreted as the
density of the mass quantum states in a small vicinity of m1. As
it is not hard to see (50) has a form almost analogous to the
Fermi golden rule. Thus, it is demonstrated that Hawking radiation
can be satisfactorily described in the suggested, simplified way.
Or, Hawking radiation can be simplifiedly considered as the result
of the decay (induced by small gravitational perturbation) of
ground standing wave in the two traveling waves.

Further, since for the Schwarzschild black hole $M=M_{g}$, then
(43), (47), (49), imply
\begin {equation}
   t_{ev}\simeq \frac {(\frac {M}{m_{1}})}{3} \tau_{1}\simeq \frac {N}{3}\tau_{1}\sim  N \tau_{1}               .
\end {equation}
or
\begin {equation}
  \frac {t_{ev}}{\tau_{1}}\simeq \frac {1}{3}\frac {M}{m_{1}}\sim \frac{M}{m_{1}} .
\end {equation}
Formally speaking black hole evaporates completely when one third,
or more roughly, all initial mass quantums decay one by one (one
after one). Even if such statement is not exactly true it is
consistent within suggested linear approximation.

\section {Discussion and conclusion}

Now we can discuss previously obtained results. Obviously, use of
the classical (or weak general relativistic) mechanical,
electrodynamical, thermodynamical and statistical principles in
common with simple quantum mechanical concepts, lead toward
suggested, simple, linear approximation of the quatum gravity.
Even if such approximation is rough, it is not useless for
description of the basic dynamical and thermodynamical
characteristics of black hole. On the contrary, basic dynamical
(horizons) and thermodynamical (Bekenstein-Hawking entropy,
Bekenstein entropy/surface quantization, Hawking temperature and
Hawking radiation) black hole characteristics can be obtained very
simply within given approximation, by use of the usual algebraic
equations only. Black hole entropy and temperature can be obtained
by ground standing wave at horizon, while Hawking radiation can be
obtained by  gravitational decay of this standing wave in its
traveling components. But, of course, many details of the
description of the black hole dynamics and thermodynamics stand
without given approximation. For this reason suggested approximate
method for the description of the basic dynamical and
thermodynamical characteristics of black hole can be very useful
for the quantum gravity non-specialists.

\section {References}

\begin {itemize}

\item [[1]] {\it General Relativity: An Einstein Centenary Survey}, S. W. Hawking, W. Israel eds. (Cambridge University Press, Cambridge, 1979)
\item [[2]] R. M. Wald, {\it General Relativity} (Chicago University Press, Chicago, 1984)
\item [[3]] R. M. Wald, {\it Quantum Field Theory in Curved Spacetime and Black Hole Thermodynamics} (Chicago University Press, Chicago, 1994)
\item [[4]] {\it Black Holes, Gravitational Radiation and the Universe}, B. R. Iyer, B. Bhawal eds. (Kluwer, Dordrecht, 1998)
\item [[5]] V. P. Frolov, D. I. Novikov, {\it Black Hole Physics: Basic Concepts and New Developments} (Kluwer, Dordrecht, 1998)
\item [[6]] L. Susskind, J. Uglum, {\it Black Hole Entropy in Canonical Quantum Gravity and Superstring Theory}, Phys. Rev. D50 (1994) 2700
\item [[7]] A. Strominger, C. Wafa, {\it Microscopic Origin of the Bekenstein-Hawking Entropy}, Phys. Lett. {\it B379} (1996) 99
\item [[8]] S. Hossenfelder, {\it What Black Holes can teach Us}, hep-ph/0412265
\item [[9]] D. V. Fursaev, {\it Can One Understand Black Hole Entropy without Knowing Much about Quantum Gravity}, gr-qc/0404038
\item [[10]] P. S. Laplace, {\it Exposition du Systeme du Monde}, Vol. II (Paris, 1796)
\item [[11]] V. Stephani, {\it La Place, Weimar, Schiller and the Birth of Black Hole Theory}, gr-qc/0304087
\item [[12]] R.P. Feynman, R. B. Leighton, M. Sands, {\it The Feynman Lectures on Physics}, Vol. 2, (Addison-Wesley Pub. Co. Inc., Reading, Mass., 1964)

\end {itemize}

\end {document}